# Foreign body reaction (immune respond) for artificial implants can be avoided

Irina Kondyurina[1], Alexey Kondyurin[2,3]

[1]School of Medicine, University of Sydney, Sydney, Australia
[2]School of Physics, University of Sydney, Sydney, Australia; aleksey.kondyurin@sydney.edu.au
[3]Ewigar Scientific, Ewingar, NSW, Australia

**Abstract:** Despite on great success with artificial implant for human body, the modern implants could not solve major health problems. The reason is an immune reaction of organism on artificial implant known as foreign body reaction. We have found a way to avoid or decrease the foreign body reaction. The surface of artificial implant is modified with condensed aromatic structures containing free radicals, which provide a covalent attachment of host proteins in native conformation. The total protein coverage prevents direct contact of immune cells with the implant surface and the immune cells are not activated. As result the immune respond of organism is not generated and the artificial implant is not isolated from the tissue: no collagen capsule, low activity of macrophages, low cell proliferation, low inflammatory activity.

## 1. Introduction

Modern artificial implants can successfully work in organisms for more then 20 years. It was estimated that only in USA 20 millions of people have a biomedical implants to save their life [1]. The success of the medical implant industry and surgery practice is proved by a high number of successful operations. However, the artificial implants are not complete solution. Research has shown that 73% of people who received artificial heart implants survived after 9 years and 65% survived after 18 years. In the case of artificial aorta implants 85% of people survive after 5 years.

The problem is a full integration of the implant into an organism. An intrusion of any artificial material into an organism causes a reaction of the organism's immune system called Foreign Body Reaction or Foreign Body Respond (FBR) [2]. The immune reaction on the foreign body protects the organism against bacteria, viruses and injuries, and causes an isolation of the implant from the organism tissue that can break a functionality of the implant. In the worst case scenario the implant must be removed or replaced. This involves a secondary operation that increases the risk of lethality significantly, especially for elderly people who mostly need the implants. Research has shown that 100% of all artificial implants cause an immune reaction, and 35% of them require a secondary operation.

The immune response on the artificial implant is manageable with immunodepressant therapy [1, 3, 4]. The immune system of entire organism is depressed to decrease or to avoid foreign body reaction on the implant. Such therapy is a common way to avoid the implant isolation. However, such depressed immune system cannot protect an organism against any disease. Most of such patient death is caused by another disease, not connected with implant itself.

The Foreign Body Reaction is ignited by the implant surface interaction with host proteins (Figure 1) [5]. The protein molecules are adsorbed on the implant surface and the initial conformation of protein is damaged [6]. The immune cells recognize the protein with damaged conformation as a foreign protein and the immune cells are activated to destroy the protein. The cycle of attachment - damage conformation - destruction is repeated and repeated up to the immune cells activate next level of the immune respond such as neutrophil and macrophages collection near the implant surface, intensive cell proliferation and angiogenesis, collagen encapsulation and calcification (in worse case) of the collagen capsule to isolate the implant [6, 7]. The Foreign Body Reaction is detected by immune cell activity such as neutrophils and macrophages, intensive cell proliferation, inflammatory reaction, specific cytokines, and collagen capsule surrounding the implant.

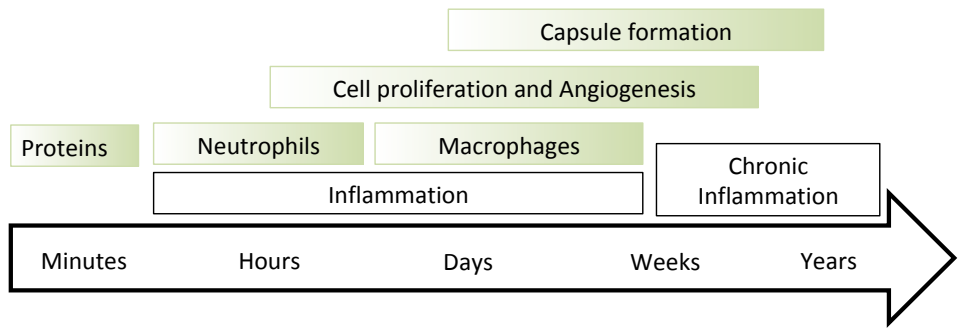

**Figure 1.** Scheme of foreign body reaction with main stages from the time of first contact of the implant with organism tissue. The scheme is adapted from literature data [5] for untreated polyurethane.

The decrease of Foreign Body Reaction was observed earlier in our studies on ion beam implanted polyurethane implants in mice and rats [8]. The collagen capsule was not formed near the carbonized polyurethane surface even after 5 months in mice organism. The low immune respond was based on protein covalent attachment on the carbonized surface due to presence of free radicals at the edge of graphene sheets [9, 10]. However, the detailed histological investigation of the tissue surrounding the implant was not performed.

In the present study we investigate the immune respond of organism on polyurethane implants treated by plasma immersion ion implantation (PIII) as a kind of general ion beam implantation method and describe a method to avoid the foreign body reaction for polyurethane implants without immunodepressant therapy.

2. Materials and Methods

Polyurethane (Figure 2) was synthesized from polypropylene glycol terminated by toluenediisocyanate (Aldrich) and polyoxitetramethylene (Aldrich) with a deficit of hydroxyl groups that gave 3D crosslinking of the macromolecules. The details of the synthesis are described in elsewhere [11]. The synthesized polyurethane was swollen in heptane to remove all non-crosslinked molecules and then dried on air up to complete evaporation of heptane controlled by FTIR spectra. The final films of 0.3 mm thickness and 150 mm diameter were used.

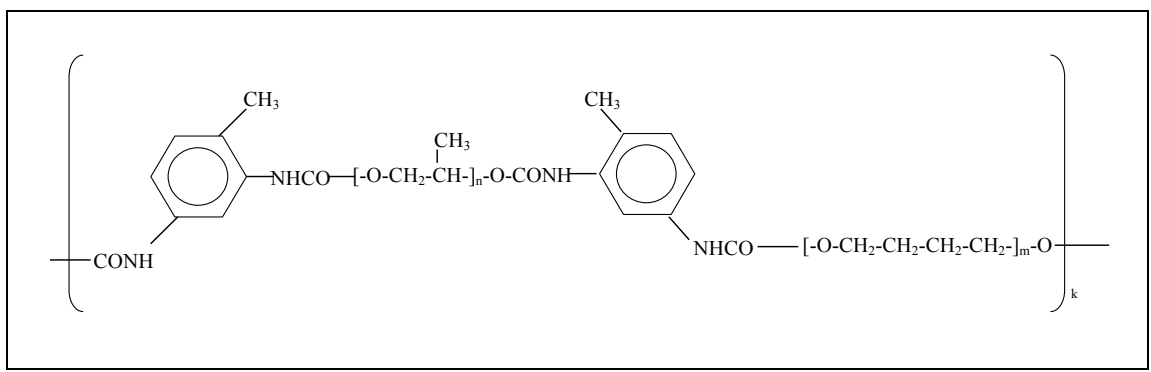

**Figure 2.** The general formula of PPG-DI-PTHF polyurethane. Crosslinking bonds are not shown.

The polyurethane films were treated by Plasma Immersion Ion Implantation (PIII) by 20 keV energy nitrogen ions of 40, 80, 200, 400 and 800 sec treatment time. The details of PIII treatment were described elsewhere [12]. Next day the films were cut to 3.5 diameter disks and implanted subcutaneously in mice. This experiment has been approved by the University of Sydney Animal Ethics Committee (protocol number K20/12-2011/3/5634) and conducted in accordance with the Australian Code of Practice for the Care and Use of Animals for Scientific Purpose. After 7 days the implants were taken from mice, sliced with microtome and investigated with Milligan's trichrome, Hematoxylin-Eosin, F4-80, Ki-67 and Von Willebrand Factor staining following standard protocols [13]. The micro-photos were done with the Nikon microscope (Japan). The objectives x4 for low resolution image and x20 for high resolution image were used.

Analysis of protein attachment on polyurethane surface was done with Bovine Serum Albumin (BSA) as an example. The polyurethane samples were incubated in BSA solution for 24 hours at room temperature. Protein was diluted to 20 μg/ml in 10 mM sodium phosphate buffer pH 7. After incubation, the samples were regressively washed in the buffer and in deionized water. For covalently attached protein analysis, the polyurethane samples were washed in 2% sodium dodecyl sulphate (SDS) solution in deionised water at $70^0$C temperature for 1 hour.

FTIR ATR spectra in a range of 400-4000 $cm^{-1}$ was recorded with the spectrometers Nicolet Magna 650 (USA), Digilab FTS (USA) and Excalibur (USA) were used. ATR spectra were recorded with ATR crystal Ge, the angle of the beam incident was 45 degrees, and number of scans was 100. The spectral resolution was 4 $cm^{-1}$. A contact angle of deionized water and diiodomethane (Aldrich, Australia) on the polyurethane surface was measured with DS-10 device (KRUSS, Germany). Micro-Raman spectra were recorded in $180^0$ degree geometry excited by Nd:YAG laser line of secondary harmonic (532.14 nm wavelength) on a diffraction double monochromator HR800 (Jobin Yvon). The electron spin resonance spectra (ESR) were recorded using Adani electron spin resonance spectrometer (Minsk, Belarus). The surface topography of the polyurethane was measured with Atomic force microscopy (AFM) using Park System device (Park System, South Korea) in tapping mode with a speed of 1 Hz and amplitude 10 nm. Microphotos of the polyurethane surface were done with using of optical microscope MBS-10 attached video camera.

## 3. Analysis of PIII treated polyurethane surface

The untreated polyurethane surface is hydrophobic with water contact angle of 93 degrees. The water drop measurements show significant decrease of the contact angle to 35-45 degrees in 15 minutes after PIII treatment. The contact angle increases with the storage time after PIII treatment up to 60-70 degrees. The angle is stabilised a week after PIII treatment. These changes were observed to be in a similar statistical range for all PIII treatment times from 40 sec to 800 sec. The contact angles of the water and diiodomethane were used for calculation of surface energy and its parts. The surface energy is an energetic parameter of the surface to describe an interaction with the adsorbed molecules. The surface energy increases from 33 mN/m for untreated polyurethane to 65-67 mN/m for freshly treated polyurethane. A week after PIII treatment the surface energy decreases with stabilisation at 45-50 mN/m. The stabilized level of the surface energy is much higher then the surface energy of untreated polyurethane. The dispersic part of the surface energy increases from 31 mN/m for untreated polyurethane to 40-44 mN/m for freshly treated polyurethane and decreases with storage time up to 37-39 mN/m. The polar energy changes more dramatically. The polar part calculations show a value of 2 mN/m for untreated polyurethane. A short time after PIII treatment the polar part of surface energy increases to 17-27 mN/m and then decreases and is stabilised at 8-14 mN/m. Therefore, the surface is hydrophilic after PIII treatment.

The Raman spectrum of PIII treated polyurethane (Figure 3a) contains new strong lines at 1348 and 1595 cm$^{-1}$. These two lines are interpreted as resonance Raman lines of the carbonised top layer: D and G peaks. G-peak is $E_{2g}$ vibrations in graphitic structure with sp$^2$ hybridisation of the valence electrons. D-peak is $A_{1g}$ vibrations in graphitic structure. The ratio I(D)/I(G)=1.77 and G-peak position corresponds to I area in Robertson/Ferrari diagram [14] that shows a nanocrystalline graphite with characteristic size of graphitic islands of $L_a$=2.5 nm separated by amorphous carbon with sp$^3$ hybridization.

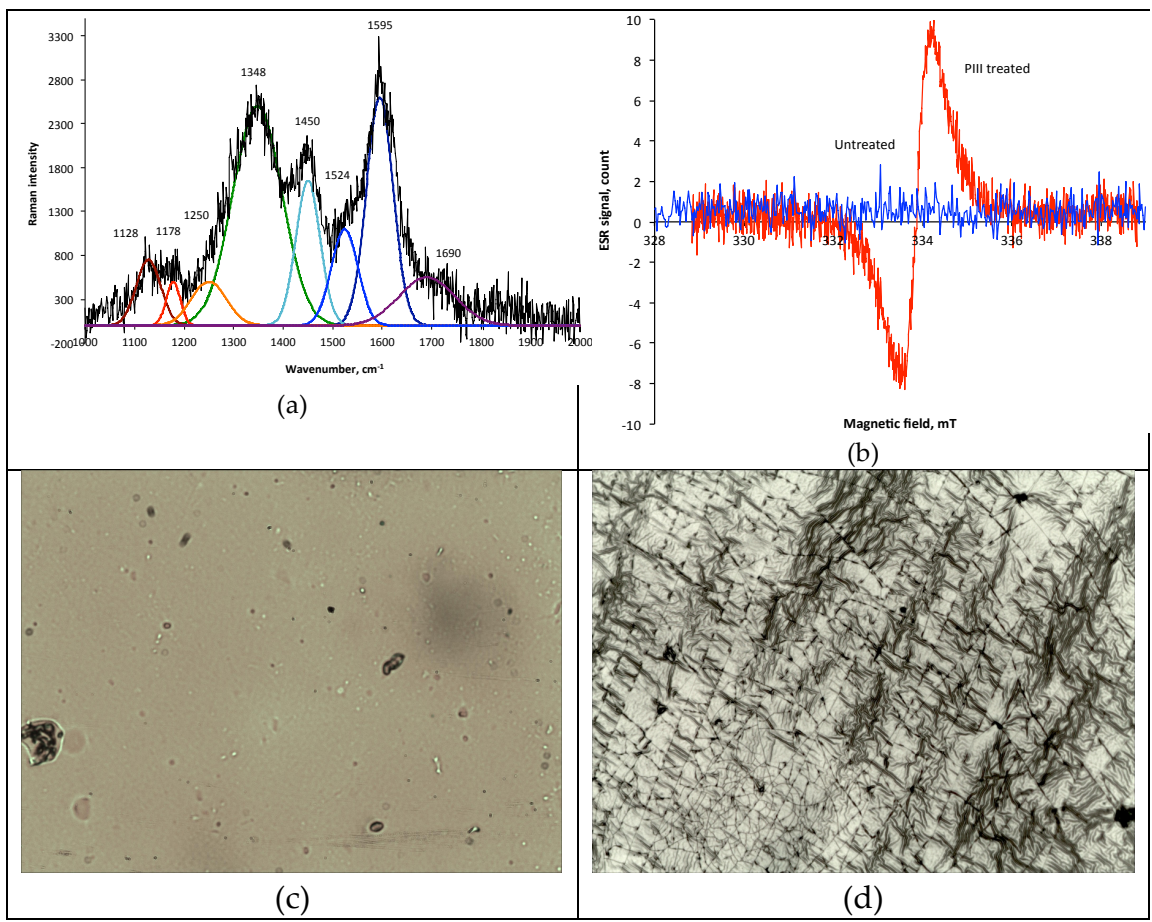

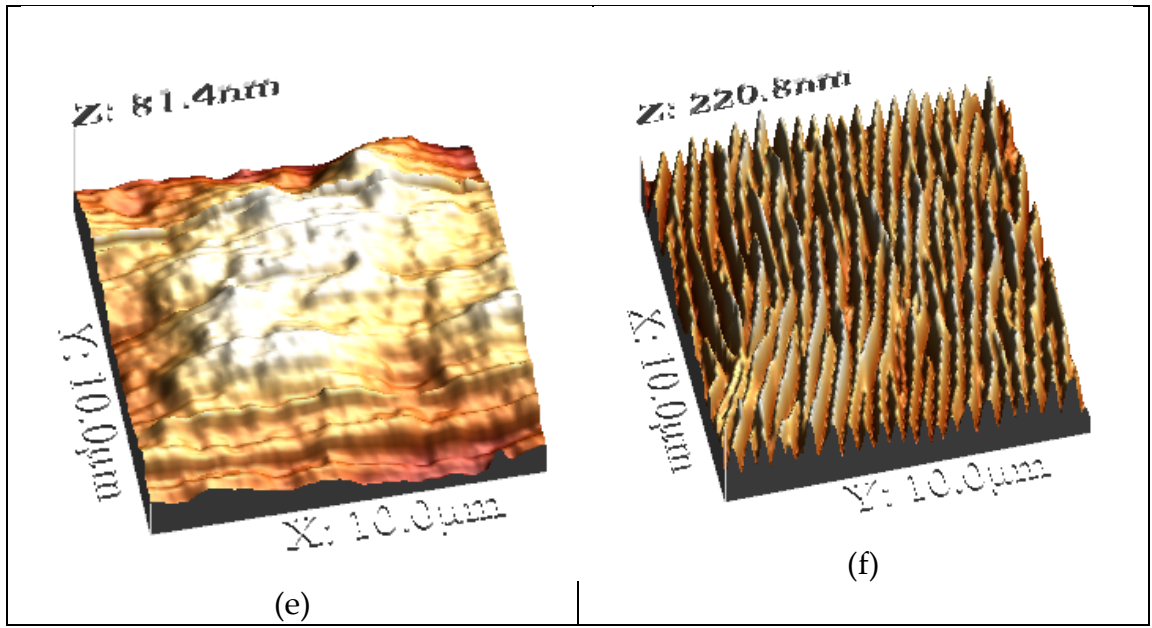

**Figure 3.** (a) Micro-Raman spectra of 20 keV nitrogen ion PIII treated polyurethane at 800 sec (bottom spectrum). Two new strong peaks at 1348 and 1595 cm-1 show the presence of graphite planes with defects in PIII treated polyurethane. (b) Electron spin resonance spectra of untreated polyurethane (blue) and PIII treated polyurethane (red) recorded from the sample after 3 months storage in laboratory. Optical microphotos of polyurethane untreated (c) and after 800 sec PIII treatment by 20 keV energy nitrogen ions (d). The PU was synthesised and treated on silicone wafer to exclude any deformations after the treatment. Objective is x50. The developed surface structure with waves and cracks is observed after PIII treatment. AFM 3D 10x10 μm area image of the polyurethane surface untreated (e) and treated with PIII for 40 sec with 20 keV nitrogen ions (f).

A signal of unpaired electrons in the modified polyurethane has detected with the electron spin resonance spectra. The electron spin resonance spectra of the untreated polyurethane and the PIII treated polyurethane at 800 seconds by nitrogen ions with 20 keV energy are presented in Figure 3b. The untreated polyurethane film does not have a signal higher than the noise level. The PIII treated film has a strong signal with g-factor about 2.0028. The signal with such g-factor corresponds to unpaired electrons of carbon free radicals at the edges of aromatic structures in graphite- or graphene-like structures [15]. The spectra show that the free radicals in the modified layer appear after PIII treatment and then remain when it is stored for a further 3 months. Such long shelf-life free radicals are stabilised by π-electrons of the aromatic rings of graphite-like structures.

The surface topography of PPG-DI-PTHF-0.35 polyurethane was changed dramatically after PIII treatment that is visible with optical microscope and Atomic Force Microscope (Figures 3c-f). The treated surface is wrinkled and cracked. The direction of the wave and cracks are random. The surface topography remained stable after a month long storage time. The topography viewed was similar to polyurethane free standing films and polyurethane coating on hard substrate (silicon wafer). Therefore the surface wrinkles and cracks are not the result of a deformation of the sample. The surface topography corresponds to hard surface layer with internal stresses over the soft bulk layer.

The Bovine serum albumin (BSA) has been used as an example in the attachment experiment. The polyurethane samples were incubated overnight in BSA solution in PBS buffer. The control samples were incubated in buffer solutions only. These were prepared the same way and at the same time as the protein attached samples. Then all samples were rinsed in deionized water and dried overnight, the FTIR ATR spectra was recorded. The spectral lines of protein are not visible in raw spectra of the samples as recorded due to high intensity of the polyurethane lines in comparison with the protein lines. The low intensity of protein lines corresponds to very thin BSA protein monolayer (about 5 nm) in comparison with the high intensity polyurethane spectrum. The spectra of control samples were subtracted from the spectra of protein samples treated at the same PIII treatment time with adjusted subtraction factor to minimise absorbance of polyurethane lines (Figure 4a). The subtracted spectra show Amide A line at 3300 cm$^{-1}$, Amide I line at 1650 cm$^{-1}$ and Amide II line at 1540 cm$^{-1}$. The line positions of Amide A and Amide II in the spectra of pure polyurethane and pure protein are similar. These lines in the subtracted spectra can be interpreted as related to both polyurethane and protein. Therefore these lines cannot be used for analysis of protein attachment. The Amide I line in the spectra of pure polyurethane was observed at 1720 cm$^{-1}$, and the Amide I line in the spectra of pure protein was observed at 1650 cm$^{-1}$. Due to these different positions, the Amide I line can be used for analysis of the protein attachment. The absorbance of Amide I line normalised on the line of polyurethane vibrations at 1100 cm$^{-1}$ as a stable intensive line is presented in Figure 4b according to PIII treatment time. The amount of attached BSA increases with PIII treatment time up to saturation level. The saturation level observed after 800 seconds of PIII treatment time is double than for the untreated polyurethane.

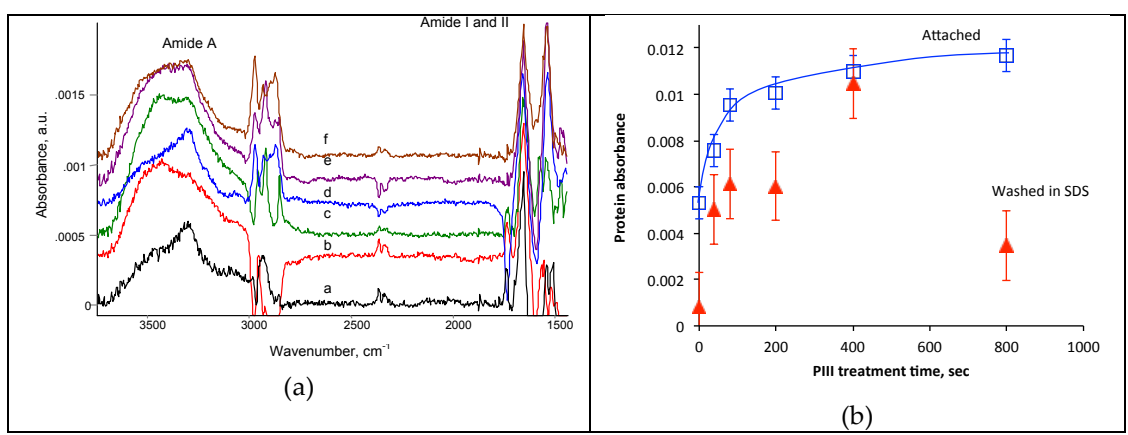

**Figure 4.** (a) FTIR ATR spectra of attached BSA on untreated and PIII treated polyurethane of PPG-DI PTHF with 0.35 NCO/OH ratio. From bottom to top: (a) untreated polyurethane, (b) 40, (c) 80, (d) 200, (e) 400, (f) 800 s PIII treatment time with 20 keV nitrogen ions. The spectra of corresponding polyurethanes are subtracted. (b) FTIR ATR spectra absorbance of Amide I line of BSA attached to polyurethane of PPG-DI PTHF with 0.35 NCO/OH ratio with PIII treatment time: blue is attached BSA, red is BSA attached and then washed off in SDS 1 hour at 70°C. The absorbance of Amide I of protein was normalized on the absorbance of 1100 cm$^{-1}$ line related to vibrations in polyurethane.

The detergent SDS was used for washing the non-covalently attached BSA molecules from the polyurethane surface. The experiment was done with the same samples of polyurethane after FTIR spectra recording. After washing in the detergent, all samples were rinsed in deionized water and dried overnight. The FTIR ATR spectra of control samples were subtracted from the spectra of protein samples as described above and absorbance of protein Amide I line was calculated and presented in Figure 4b. The protein lines in the spectrum of untreated polyurethane are not observed, and are under the noise level of the spectrum. The spectra of SDS washed samples show the Amide I line of the attached protein only for PIII treated polyurethane. Therefore the protein was almost washed from the untreated polyurethane. It shows that the protein on untreated polyurethane attached due to physical forces was disturbed by the physical forces of the detergent's molecules. Washing in detergent was sufficient to remove whole protein layers attached to the untreated surface due to physical forces.

The amount of attached protein on PIII treated polyurethane after washing in detergent solution is lower than the amount of attached protein before the washing. However, a significant portion of the attached protein remains on the surface after washing despite the detergent's physical forces. This shows that this protein is attached chemically to the surface. The ratio of attached to remained after detergent washing varies from 30% for 800 sec PIII treated sample to 95% for 400 sec PIII treated sample with average amount 63% for all PIII treated samples.

**4. Histology analysis of the tissue around the polyurethane implant**

The histological images of the sections stained with Milligan' trichrome show that a collagen capsule is formed around all samples of the polyurethane implants (Figure 5). The polyurethane implants were removed at the microtome cutting due to very elastic polyurethane properties in comparison with the surrounding tissues. In all the images, the polyurethane was in the centre of the empty ovals. In some cases, the residuals of the polyurethane implants remained in the sections and stained slightly. The capsule continuously covers the implants, isolating the implant from the tissue. The observed ruptures in some capsules are due to the microtome cutting as shown by the character of the ruptures. The capsule for untreated polyurethane implants had more intensive colouring and was broader than for PIII modified implants. However, the capsule thickness appeared to be independent of what surrounded the implants, whether muscle or connectivity tissues.

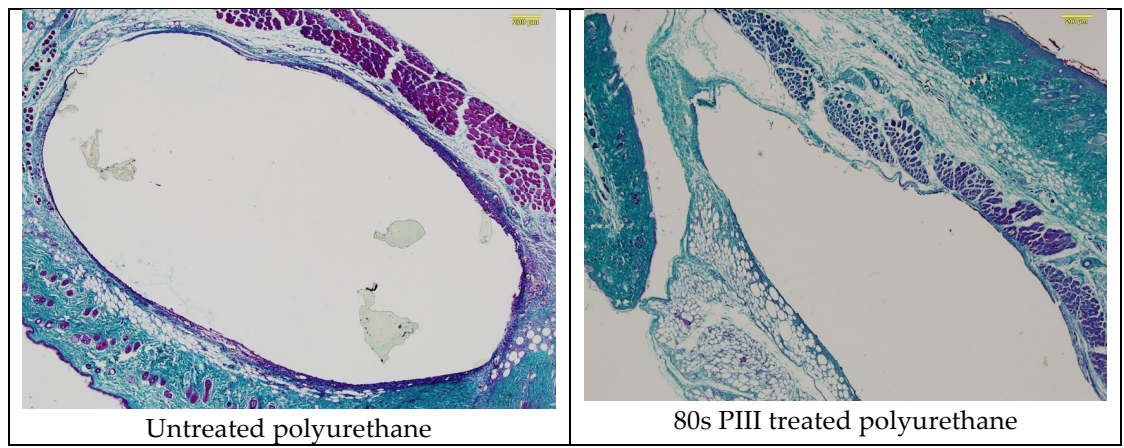

| Untreated polyurethane | 80s PIII treated polyurethane |

**Figure 5.** Images of the histological samples of polyurethanes stained in Milligan. Objective x4. The implant position was in the centre of the ovals. Usually, the polyurethane sample was removed at the microtome cutting, but sometimes the small pieces of the polyurethane samples remain.

The thickness of the capsule was analyzed quantitatively using the high resolution images of the tissue near the implants. The representative examples of the section stained with Milligan' trichrome in high resolution for polyurethane implants with different PIII treatment times are presented in Figure 6.

The collagen shell stained Milligan's trichrome was well developed in the tissue surrounding the untreated polyurethane implants. The collagen fibres were structured and directed parallel to the implant surface on a distance of 40-60 μm. This part of the shell was characterised by dense collagen structure. The collagen fibres were disordered at a distance of

60-120 μm from the implant surface. This part of the shell was less dense but denser than the collagen and elastin fibre structures in the normal tissue. In some samples the very dense collagen structure was observed in the first approximately 10 μm layer from the implant surface.

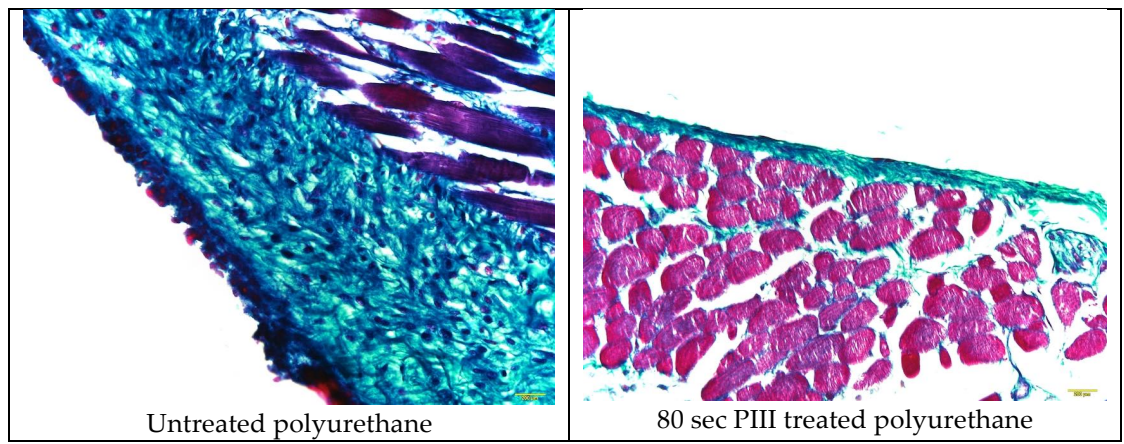

**Figure 6.** Microphotographs of the histological tissue samples stained in Milligan for untreated and PIII treated polyurethane. An empty space on the figures is the implant position.

The PIII treatment of the polyurethane implant changed the collagen shell in the organism (Figure 6). The dense collagen shell was not observed. The collagen shell near the implant surface was not developed and the collagen fibres were not directed but were disordered and rare. The whole collagen shell was thinner than for the untreated polyurethane implant. For some samples the shell was observed in 1-2 cell layers between the implant surface and the muscle cells. In some samples the dense collagen shell was observed in the thin surface layer (up to about 10 μm), which is in closest contact with the implant. All these difference in the collagen shell were observed in different animals and are not attributed to a specific immune reaction of one mouse.

The average thickness of the collagen shell was calculated and is presented in Table 1. The staining of all the slides was done in one day, the microphotographs were done in one microscope in one day with adjusted intensity and optical elements of the microscope, and the analysis of the images was done for all microphotographs simultaneously. This allowed the results between the different samples in the whole batch to be compared quantitatively.

The results show that the average thickness of the collagen capsule overall PIII treated samples was significantly less for the PIII treated polyurethane implants (41 μm) than for untreated implants (102 μm) ($p<0.001$). The significant decrease in the thickness from 102 μm to 46 μm was observed after a short 40 second PIII treatment time ($p<0.01$). This difference between the untreated and PIII treated samples increases with the PIII treatment time. The thinnest thickness (38 μm) is observed for 200 sec PIII treatment time.

Table 1. Thickness (μm) of the capsule near the polyurethane implants in dependence on PIII treatment time.

|  | Untreated | 40 sec | 80 sec | 200 sec | 400 sec | 800 sec |
|---|---|---|---|---|---|---|
| Thickness, μm | 102±21 | 46±23 | 42±22 | 38±14 | 40±6 | 37±11 |

The microphotographs of H&E stained tissue are shown in Figure 7. The tissue near the untreated polyurethane implant contains macrophages, monocytes and fibroblasts. Multi-nuclear foreign body giant cells were not observed. The macrophages and monocytes were mostly distributed near the surface of the implant in the capsule. Their concentration was highest in the capsule near the implant surface. The fibroblasts were positioned mostly on a distance from the implant surface, where the collagen fibres dominate. There was no cell lysis or necrosis.

The tissue near the PIII treated implant is less disturbed (Figure 7). Most cells in the capsule are fibroblasts and few cells were found in some spots like macrophages and monocytes (marked with arrows). The macrophages and monocytes were predominantly in the contact zone of the implant surface. In some images it is clear that the spots with macrophages and monocytes are distributed along the implant surface at the same periodicity as the cracks and waves of the polyurethane surface. Such cell distribution was observed for all PIII treatment times. Similar capsule images were observed in cases when the modified implant was in contact with the connective tissue or muscle tissue or adipocytes cells. In some places where the tissue contacted the modified implant the capsule was not observed at all. The cells, which were in contact with the implant surface, were fibroblasts or the connective tissue fibres.

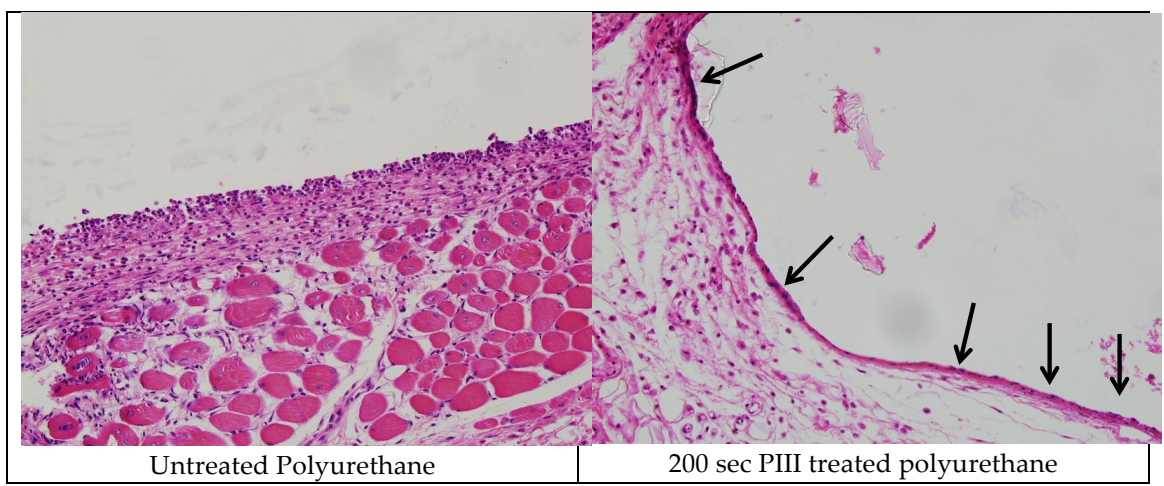

| Untreated Polyurethane | 200 sec PIII treated polyurethane |

**Figure 7.** Microphotographs of the histological tissue samples stained in Hematoxylin-Eosin for the untreated and 200 sec PIII treated polyurethane implants. An empty space on the figures is the implant position. Positions of macrophages are shown with arrows.

The staining with F4-80 antibody was used to detect a presence of macrophages in the tissue surrounding the polyurethane implants. The high macrophages activity was observed in the collagen capsule near the surface of the untreated polyurethane implant (Figure 8).

The rest of the tissue outside of the collagen capsule showed low macrophages activity. The same low macrophages activity was observed in the tissue surrounding the PIII treated implant (Figure 8). For some samples macrophages activity near the implant surface was not observed.

The integral area of the tissue with brown colour and the integral intensity of the colour was analysed to estimate the macrophages activity in the tissue. The results of activity macrophages dependence on the PIII treatment time are presented in Table 2. The highest macrophages activity was observed in the tissue surrounding the untreated polyurethane implant and 800 second PIII treated sample. The lowest macrophages activity was observed for 200 second PIII treated implants. The difference between the untreated and 200 second PIII treated samples is significant ($p<0.001$).

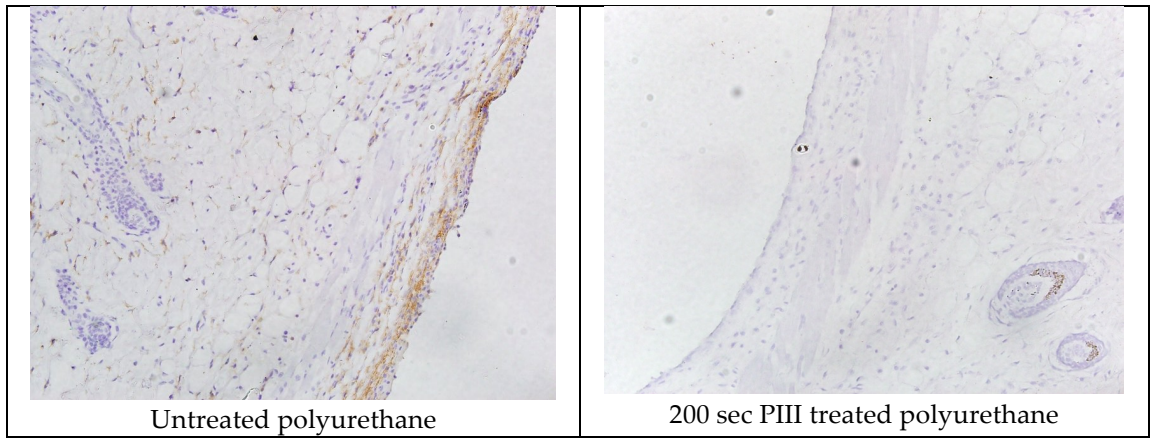

**Figure 8.** Microphotographs of the histological tissue samples stained in F4-80 antibody for the PPG-DI-PTHF-0.35 polyurethane implants. An empty space on the figures is the implant position.

Table 2. Macrophages activity near the polyurethane implants in dependence on PIII treatment time.

|  | Untreated | 80 sec | 200 sec | 400 sec | 800 sec |
|---|---|---|---|---|---|
| Area, % | 3.6±1.0 | 1.6±0.6 | 0.6±0.3 | 1.3±0.4 | 4.4±1.1 |
| Density, a.u. | 9.8±2.7 | 4.0±1.4 | 1.7±0.8 | 3.4±1.1 | 13.7±4.0 |

The Ki-67 antibody was used for detection of the cell proliferation activity in the tissue surrounding the polyurethane implants. The example of characteristic images for untreated and PIII treated implants with different PIII treatment time are presented in Figure 9. The colour distribution and intensity shows a high cell proliferation activity in the tissue near the untreated polyurethane implants. The maximum activity is observed in the collagen capsule and the activity gradually decreases with the distance from the implant.

The cell proliferation activity near PIII treated implants was much less (Figure 8). In some samples the activity was observed only in a thin surface layer corresponding to the one cell thickness. Most intensive activity was observed in fields of new vessel formation.

For a quantitative analysis of the cell proliferation activity the colour area and density of the Ki-67 staining was analysed and is presented in Table 3. The activity decreases with PIII treatment time and gets a minimal value at 400 seconds of the treatment time (p<0.001). This trend was observed in the area and the density of stained cells.

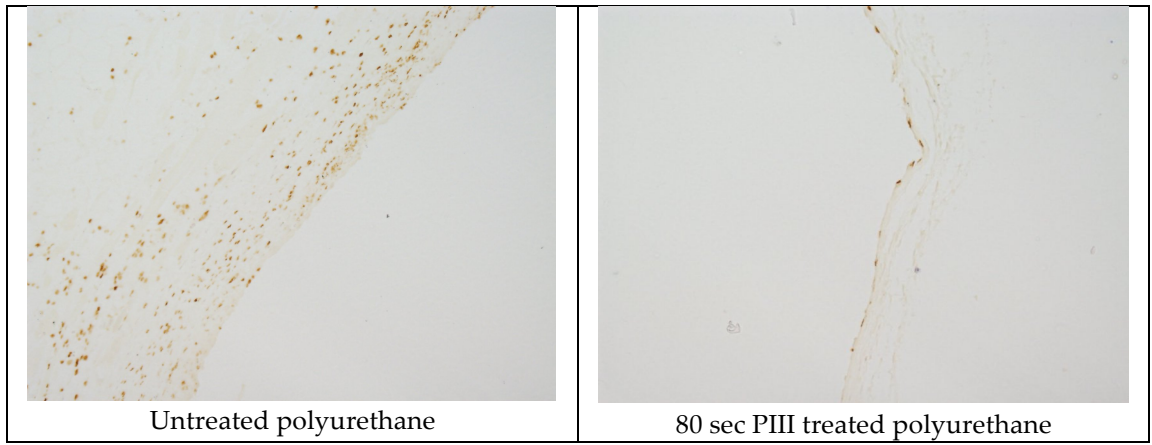

**Figure 9.** Microphotographs of the histological tissue samples stained in Ki-67 antibody for untreated and 80s PIII treated polyurethane. An empty space on the figures is the implant position.

Table 3. Cell proliferation activity near the polyurethane implants in dependence on PIII treatment time.

|  | Untreated | 40 sec | 80 sec | 200 sec | 400 sec | 800 sec |
|---|---|---|---|---|---|---|
| Area, % | 1.12±0.37 | 0.88±0.26 | 0.51±0.30 | 0.47±0.25 | 0.32±0.11 | 0.37±0.23 |
| Density, a.u. | 4.4±1.5 | 3.3±0.7 | 1.9±1.3 | 1.9±0.9 | 1.2±0.4 | 1.4±0.8 |

Von Willebrand Factor (vWF) as a proinflammatory protein and a key player in hemostasis was analysed in the tissue surrounding the polyurethane implants. The microphotographs of the tissue stained in the vWF antibody are presented in Figure 10. The tissues near the untreated polyurethane implants accumulated a large amount of the vWF. Some samples show the highest concentration of the vWF in the capsule, but some samples show a distribution of the vWF in the whole tissue far from the capsule (an example on Figure 10). The vWF high amount was also observed in the new vessels. In PIII treated samples the vWF was concentrated in spots outside the capsule. Much lower amount vWF was observed in the tissue surrounding the PIII treated polyurethane implants. In some samples, the coloured area was not observed at all. In all these samples the capsule did not contain vWF.

For quantitative measurements the area of vWF stained tissue was calculated (Table 4). The measurements show that the tissue near the implants of 80-800 s range of PIII treatment time has a significantly lower area of vWF than the tissue surrounding the untreated and 40 s PIII treated implants (p<0.001). The difference between the tissues near the PIII treated implants in 80 second - 800 second range was not significant (p>0.05).

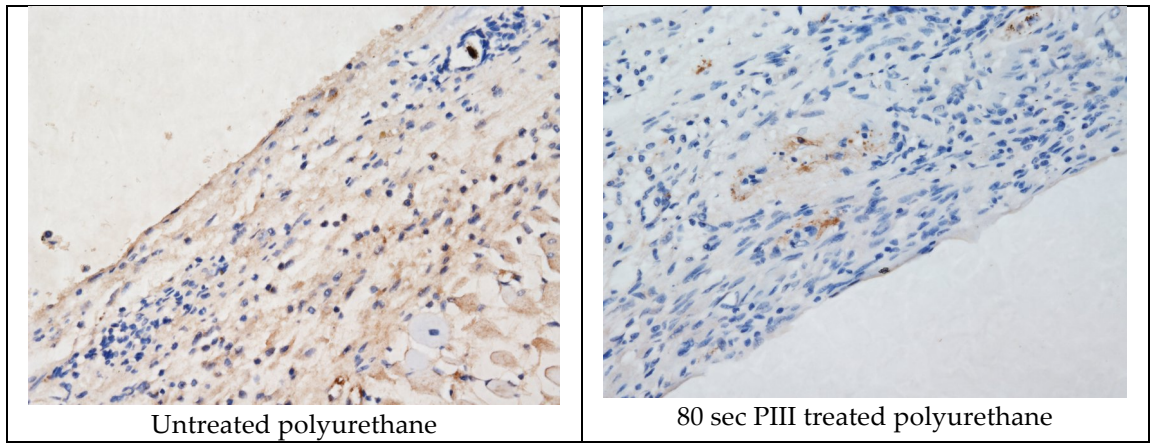

**Figure 10.** Microphotographs of the histological tissue samples stained in Von Willebrand Factor for untreated and 80 sec PIII treated polyurethane. An empty space on the figures is the implant position.

Table 4. Inflammatory activity near the polyurethane implants in dependence on PIII treatment time.

|        | Untreated | 40 sec | 80 sec | 200 sec | 400 sec | 800 sec |
|--------|-----------|--------|--------|---------|---------|---------|
| Area, %| 0.93±0.12 | 0.79±0.06 | 0.24±0.10 | 0.35±0.14 | 0.27±0.11 | 0.27±0.11 |

## 5. Discussion

Foreign Body Reaction (FBR) is a key problem for modern artificial implants [6]. It does not only limit the implant development but it limits the number of patients to whom the implants can be applied. An absence or reduction in the FBR would significantly improve the implant's capacity to save life. However, despite how all modern implants satisfy physical and chemical stability, non-immunogenicity and non-toxicity requirements [16], all these implants still cause an immune response in an organism. This is why there is a widely held opinion that FBR cannot be avoided [5]: "All materials implanted into humans and animals elicit 'foreign body' interactions, surrounding the materials with a protective capsule."

Attempts to avoid or decrease the immune reaction on artificial implants can be found in the literature [17-19]. All these attempts were based on stopping the immune reaction at some stage. For example, systemic corticosteroids decreases immune function and fibrosis, in particular it decreases the number of myofibroblasts [20]. However, our knowledge of the mechanism of the immune reaction is limited. The immune reaction is complicated and self-regulated with multiple ways and steps. Blocking one step in the immune response will not prevent the protective capsule formation.

Therefore, the most effective way to avoid the foreign body reaction is to do not ignite it. This way is used in the present study.

The untreated polyurethane implant shows a minor foreign body reaction with formation of the capsule filled with macrophages and fibroblasts. The giant cells or necrosis were not observed in any samples. The thickness of the capsule of about 100 μm surrounding the implant corresponded to the literature data for biocompatible implants used in medical practice [4, 21]. Therefore, the untreated polyurethane corresponds to the biocompatible materials following the organism reaction.

The organism response to PIII treated polyurethane was significantly reduced, showing that the average thickness of the capsule was significantly less near the treated polyurethane than near the untreated polyurethane. In some cases the capsule was not formed near the PIII treated polyurethane surface at all. In addition separate macrophages were observed on the surface with large distances between them whereas in other cases only normal tissue was observed. In some cases only one irregular layer of fibroblasts with thin not-structured collagen fibres was observed.

Acute inflammation and capsule formation are usually observed 5-7 days after surgery. Chronic inflammation is observed following this period with a formation of thick isolating capsule around the implant. However, if the capsule around PIII treated polyurethane is not observed 7 days from the operation, it is possible that the capsule might be not formed later because a delayed formation of the capsule is less likely. Therefore, there are cases when the capsule is not formed or has significantly less thickness with a significantly lower amount of immune cells.

For discussion of the histochemistry results this study considered a general scheme of immune reaction on an implant (Figure 1) [5] where the reaction is characterised by stages of the organism response, which becomes more complicated and intensive with time up to the total isolation of the implant from the body.

The histochemistry results showed the significant difference in immune response of the mice organism on untreated and PIII treated polyurethane implants. At first, Von Willebrand Factor concentration in the tissue near the PIII treated implants differed from that near the untreated implants. The literature [22.] demonstrates that the Von Willebrand Factor is closely related to recruitment of the neutrophils in the injured tissue near the implant. When the Von Willebrand Factor concentration increases, the neutrophils are activated and collected in the injured tissue. The images of Von Willebrand Factor assay in the tissue near the untreated polyurethane implant correspond to a moderate immune reaction on the implants as seen in the literature [23, 24]. The lower Von Willebrand Factor concentration in the tissue near the PIII treated implant means that there is a less intensive inflammation and lower concentration and activity of neutrophils. Therefore the second stage has lower intensity for PIII treated implant than for the untreated implant.

The macrophages are recruited at the next stage of the immune reaction. The role of macrophages is to remove the foreign body and/or to utilise the cell apoptosis products. The macrophages are recruited and activated via expression of cytokines released by neutrophils when the foreign body could not be destroyed in the body. The concentration and activity of the macrophages near the untreated implant show a moderate reaction of an organism consistent with the literature data [6, *25-27*]. The lower concentration and activity of macrophages in the tissue near the PIII treated implant means that the macrophages are recruited and activated with much fewer signals from the neutrophils. Therefore the third stage of the immune response has a lower intensity.

The new vessels formation and cell proliferation at the next stage provide a transport of immune cells to the injured tissue. These processes are ignited by released cytokines from the immune cells, which could not exclude the foreign body. At this stage the immune response includes deep transformation of surrounding tissues, which become involved into the healing process. The organism prepares to intensify the immune attack. The cell proliferation process in the tissue surrounding the untreated implant is much more intensive than in the tissue far away from the implant. The intensity at this stage is consistent with the literature on other implanted materials [6, 25-27]. However, the cell proliferation activity in the tissue surrounding PIII treated implant is much less. The cell proliferation is not detected, suggesting that the organism does not recognise the implant as a foreign body.

With time the implant is encapsulated with collagen fibres. The fibroblasts are activated and build a cross-linked collagen capsule around the implant. At this stage the organism tries to isolate the foreign body from the tissues. The immune response is transformed to chronic inflammation. The thickness of the capsule depends on the intensity of the foreign body reaction. The capsule around the untreated polyurethane implant is about 100 μm, which corresponds to the moderate immune reaction for most of polymer implants [4, 21].

The capsule surrounding the PIII treated implants is much thinner. This collagen capsule consists of rare disordered fibres whereas a capsule of dense ordered collagen fibres surrounds the untreated implant. In some cases the capsule near the PIII treated implant was not observed at all. Therefore, the host does not recognise the PIII treated implant as a foreign body, which must be isolated.

All these observations are related to acute inflammation observed in the 7 days after the operation. According to the literature [4, 21], capsule thickness for untreated implants grown after the acute stage, increases during the chronic stage. In addition, previous long-term investigations with ion beam treated hard polyurethane showed the absence of immune response in the organism after 5-6 months. Because of this absence in these long-term experiments with the new polyurethane, we predict that the intensive chronic inflammatory reaction on PIII treated polyurethane is unlikely due to the low inflammatory reaction during the acute stage. Therefore, the low and absent foreign body reaction is achievable.

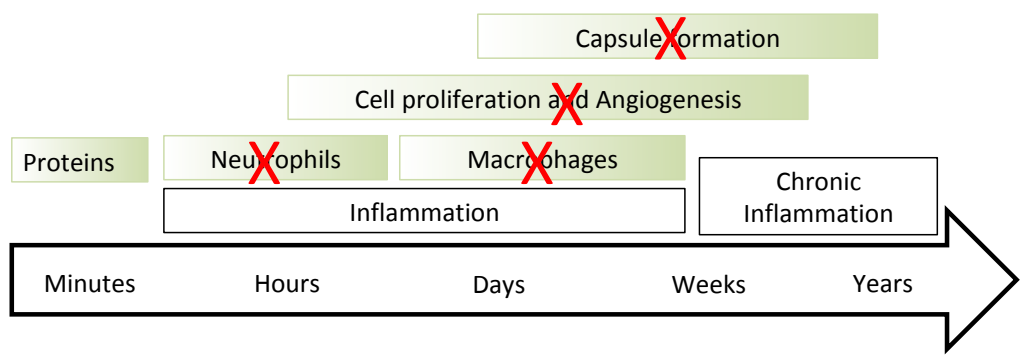

**Figure 11.** Scheme of foreign body reaction with main stages from the time of first contact of the implant with organism tissue. The scheme is adapted from literature data [5] for PIII treated polyurethane. Red crosses show low or absent activity observed in the experiments with PIII treated implants.

The thin capsule on ion beam treated implant is observed from the implant side [8]. The collagen capsule is well attached to the untreated polyurethane implant. However, the capsule is not found on ion beam implanted polyurethane surfaces with the carbonised top surface. The collagen structures are observed in the bottom of the crack of the carbonised layer, where the patches of untreated polyurethane are accessible by host proteins, phagocytes and fibroblasts.

A similar weak immune reaction was observed for ion beam treated polyurethanes in the experiments of Suzuki and colleagues [28]. Polyurethane samples treated by Ar+ were implanted into a rabbit for 16 days and a weak intensity of inflammatory reaction was observed. A similar effect of biocompatibility of polymer materials after the ion beam treatment was observed in vivo experiments with ePTFE grafts. ePTFE graft was treated by an ion beam and implanted in dogs' femoral artery. The control sample showed thrombosis after 3 days, where the modified was clear after 180 days [29]. However, the mechanism responsible for the weak immune reaction was not discovered at that time.

In the present study the attempt to prevent the phagocyte activation is based on the knowledge of specific adsorption of the host proteins on the PIII treated implant surface.

It is clear now that the immune response of an organism starts from a specific adsorption of host proteins on the surface of the artificial implant [30]. Then the phagocytes recognise the adsorbed proteins on the implant surface and become activated. The activated phagocytes release specific cytokines, chemokines, interleukins and other factors, which ignite the next steps of the inflammation reaction. The activation of phagocytes is a complex process and includes local cytokine expression, transformation of the macrophages and giant cells. At some stage the fibroblasts are activated, proliferated and ignited to produce collagen to form a capsule around the implant.

The host proteins in organisms exist in hydrophilic environment such as cytoplasm, extracellular liquids, blood and lymph. The biologically active conformation of the protein molecule is optimised for such an environment. When the host proteins come in contact with a surface of the artificial implant, the environment for the host proteins is changed: the protein is in contact with the implant, not with water molecules [31]. The protein molecule is adsorbed on the implant surface and changes the conformation due to different intermolecular interactions with the implant surface in comparison with the organism (hydrophilic) environment [26]. The conformation of absorbed protein depends on the surface properties [32]. However, all artificial implants have a surface energy in a range of 20-45 $MJ/m^2$ corresponding to the surface energy of the implant material [33]. This is much lower than the water surface energy of 72 $MJ/m^2$. Therefore, the protein conformation changes to adjust the hydrophobic interactions with the implant.

In contradiction with the untreated polyurethane, the PIII treated polyurethane implant is highly hydrophilic. The contact angle of the carbonised surface is 35-45 degrees for freshly treated polyurethane. The surface energy of the treated polyurethane 10 minutes after PIII treatment is 60-65 $MJ/m^2$. Such a high surface energy and hydrophilicity is enough to absorb the water molecules and keep the hydrophilic intermolecular interactions with proteins in such a way that the protein is surrounded by water molecules and does not change the initial conformation at the point of absorption. This effect of hydrophilicity was observed for other polymers after PIII, ion beam treatment and plasma treatment [34]. For example, the literature shows a surface energy of 71 $MJ/m^2$ for the freshly PIII treated Nylon [35] and 77.9 $MJ/m^2$ for the freshly treated polyvinyl chloride (PVC) [36]. It maintains the conformation of the absorbed protein and its activity. For example, the conformation of Horseradish peroxidase (HRP) attached to PIII treated polyethylene remains unchanged or close to the biologically active conformation as observed by Amide I line in the FTIR ATR spectra [37]. The enzyme activity of the HRP attached on PIII treated Teflon remains high [38].

At the same time protein is attached covalently to the PIII treated polyurethane implant. The covalent attachment is key to the strong anchoring of the protein to the implant surface. The protein attachment to the untreated implant can be provided with different kinds of intermolecular interactions. The van-der-Waals forces, dispersic interactions, dipole-dipole interactions, ionic interactions or hydrogen bonds can hold the protein on the surface. However, these interactions are flexible and can be easily replaced with water-protein interactions. The more hydrophilic the surface, the easier the protein can be washed with detergent or just a buffer [39]. Such a weakness of the intermolecular interaction is a reason for the Vroman effect where the proteins can be easily replaced with other proteins in the protein mixture solutions or in an organism [31]. Such a flexibility of the protein adsorption can lead to unpredictable attachment of proteins, including signal proteins for igniting the immune reaction.

The covalent attachment of protein is available with artificial chemical linkers [40-42]. The linker molecule is designed in such a way that one reactive group fixes the linker molecule on the surface and the second reactive group anchors the protein molecule to the linker. However, the chemistry of linker molecules is complicated. In most cases the linker molecules are toxic and cannot be used in an organism. Also the effectiveness of the linkers to bind the protein is much less than the total coverage. In such cases, part of the surface is not covered and the immune system can recognise the surface as foreign body.

The PIII treated polyurethane surface can be covered completely. The protein monolayer forms a total coverage (carpet) on the PIII treated polymer surface [43] including polyurethane. The protein layer covalently attached to the PIII treated polyurethane implant is not washable. The protein molecules covalently attached to the PIII treated surface cannot be replaced with other proteins and the Vroman effect is not observed [44].

The covalent attachment of protein on PIII treated polyurethane is provided via the reactions of free radicals which are formed when the polyurethane macromolecule is destroyed under high energy ion bombardment with formation of condensed aromatic structures like graphene or graphite. The free radicals are observed in the polyurethane implant after a long period of storage. The free radicals on the edges of the graphitic planes can be stabilised by π-electrons of the aromatic structure [9, 45] and remain active before the protein molecule approaches the modified surface. The literature shows that the reaction of free radicals appeared in polyethylene after ion beam treatment with aminoacids [46]. When the free radicals react with the protein molecule a covalent bond is formed [37]. As a result, the protein molecule bonds to the carbonised surface layer of the PIII treated polyurethane implant.

The molecules with free radicals in solution are widely considered to accelerate aging of an organism because the host protein molecules can be damaged in free radical reactions [47]. The result of the protein damaging and oxidation can accelerate cell lysis and necrosis. These effects were not observed in surrounding implant tissue. The effect of free radicals on the protein requires the free radicals to be mobile in a solution. The free radicals in the polyurethane implant are associated with the condensed aromatic structures which are stable in polyurethane, and cannot propagate in the solution. When the protein is bonded with the surface, the protein does not migrate into the solution. The damaging of protein or decay of the protein activity was also not observed. Therefore, the free radicals in the polyurethane surface do not influence the organism's tissues.

The endothelialisation of the implant surface is one of the important factors of the biocompatibility [48]. Endothelialisation is considered a method of reducing the immune response [49]. The attachment of active protein can significantly improve cell adhesion [50]. A fast and total endothelialisation in an organism is the primary aim for the biocompatibility of implants.

In the case of the newly synthesised polyurethane, the endothelial cells are spread and proliferate on the surface without PIII treatment. The endothelial cells also grow well on the PIII treated polyurethane surface. However, the untreated polyurethane is observed totally covered with the endothelial cells only after 6 days, while the PIII treated polyurethane is totally covered after 2 days as observed on the control tissue culture plastic (TCP) [51].

Similar cell attachment and proliferation was observed for other ion beam and PIII treated polymers. For example, the ion beam treatment of polystyrene by Na and Ne ions of 150 keV energy improved the endothelial cell spreading, attachment, proliferation and resistance to detaching with tripsin [52]. However, this research did not explain the mechanism by which this occurred. The adhesion and proliferation of endothelial cells and astrocytes on polyether sulphone and polyurethane by ion beam due to carbonisation of the polymer surface was reported in [53]. The endothelial cell adhesion and proliferation on PIII treated polyurethane was reported in [54]. The mechanism of improved cell adhesion via protein attachment due to free radical reactions was proposed in [10]. Thus, the attachment of protein with biologically active conformation leads to faster endothelialisation of the PIII treated polyurethane implant.

The observed improvement of the polyurethane implant with PIII treatment can be considered for future soft tissue implants such as breast implants, testicular implants, cardiovascular implants, vascular grafts, diaphragms, finger joints, ear implants, nose implants and others. The mechanism for reducing the immune response based on free radical reaction is critical for successful implantation.

First, the free radical concentration is reduced with time after PIII treatment. Therefore, the best result for implantation is expected when the implant is treated by plasma and installed immediately. Therefore, implants should be treated in an operating theatre just before surgery.

Second, the free radicals are sensitive to any surface contamination after the PIII treatment. Specific environmental conditions and limited contact with any other materials and devices are required before implantation into an organism.

Third, the sensitivity of free radicals limits sterilisation methods. Sterilisation by chemicals or gases reduces the concentration of free radicals and consequently the success of the implantation. Therefore, the sterilisation of the implant would be preferred before the PIII treatment. Taking into account that a plasma method is used for sterilisation [55], the PIII treatment can be done in sterile environments such as an operating theatre.

In general, the potential of implants without FBR cannot be overestimated. Cardiovascular implants including heart valve, blood vessel, catheter, pacemaker and defibrillator which do not generate FBR can be applied without immunosuppressants. Patients would not be at a risk of infection after implantation. A category of patients, who are at risk of corresponding diseases and for whom the implantation has not been recommended, will be able to be treated with the implants. The risk of inflammation and thrombosis can be significantly reduced. The breast, lips and other cosmetic implants would not cause contracture. Implants with electrical connection to the tissue such as cochlear, spine neurostimulator, implantable brain stimulator, different sensors in organs and blood would never lose electrical conductivity through the fibrotic capsule in an organism (Fig.12). The implantable drug release devices would never lose the drug penetration rate through the fibrotic capsule. Only insulin permanent releasing devices will save millions of patients with diabetes. Consequently, these improvements affect some millions of patients and can significantly influence our health, medical treatment and medical industry.

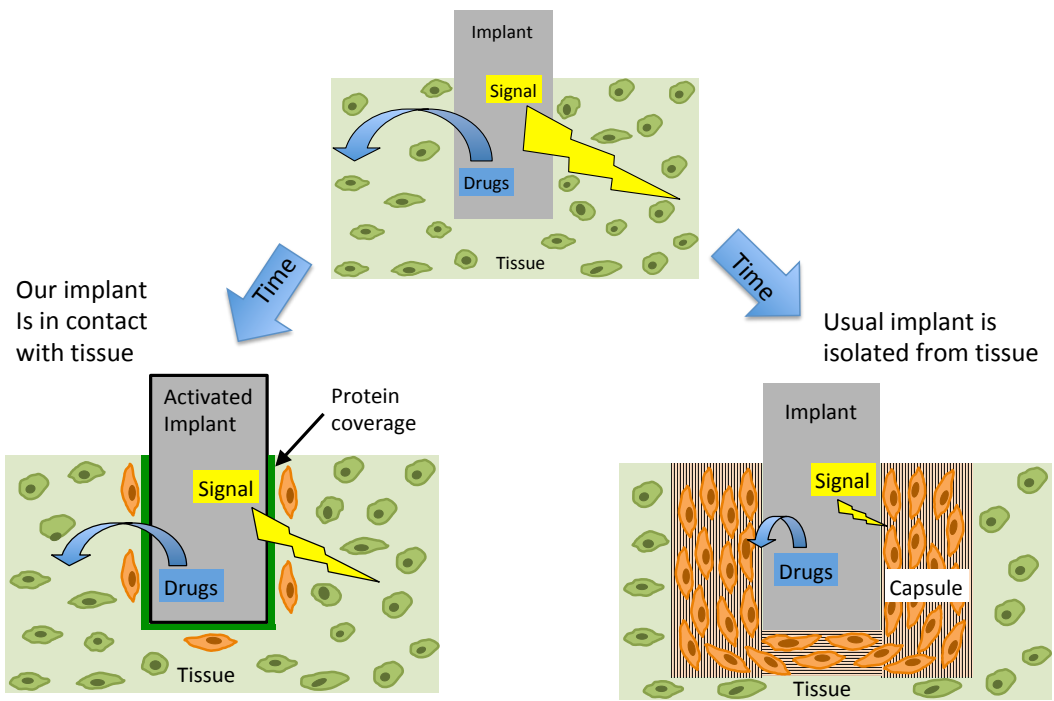

**Figure 12.** Scheme of implant with drug release or electrical connection. The foreign body reaction leads to an isolation of the implant with drug release or electrical signal. The drug release rate drops, the electrical current drops, the medical devise stops working (right image). The activated implant with the protein coverage keeps working: the drug release rate is stable; the electrical current is stable (left).

## 5. Conclusions

The investigations of the organism response on the untreated and PIII treated polyurethane implant show that all implants do not cause physiological dysfunctions in the mice. The PIII treated implants in mice show a weaker immune response of the organism, in particular:

1. The capsule around the PIII treated implant is significantly thinner ($p<0.01$).

2. The macrophages activity near the PIII treated implants is significantly weaker ($p<0.001$).

3. The cell proliferation activity near the PIII treated implants is significantly weaker ($p<0.001$).

4. The pro-inflammatory activity by vWF test is much weaker near the PIII treated implants ($p<0.001$).

Therefore, the changes in the organism response on PIII treated implant in comparison with the untreated polyurethane implant are statistically significant.

It was found that in some cases of the surface treated polyurethane the immune respond is neglectable. The reason of the low immune respond is covalent protein attachment on the implant surface. The protein attachment is provided by chemical reaction of the protein molecule with carbon atom with unpaired electron (free radical) on the edge of condensed aromatic structures with π-electron stabilizing cloud. The same unpaired electrons at the carbon atom on the surface are responsible for hydrophilic interaction with the attached protein providing saving protein conformation and activity.

Such implants with weak or without Foreign Body Reaction can potentially be applied without suppressor of immune system. We are looking for collaborators and funding to continue these studies.

**Acknowledgments:** The Raman spectra were recorded in Vibrational Spectroscopy Core Facility of University of Sydney. Authors thank Prof. B. Bao for assistance with mice surgery, Ms. Sanaz Maleki for assistance with preparation of histology samples, Prof. Marcela Bilek and Prof. David McKenzie for providing employment in an equipped plasma laboratory, Prof. Wojciech Chrzanowski for possibility to measure AFM images, Mr. V. Chudinov for recording ESR spectra. Author thanks Ms. H. Katzen and Mr. T. Kondyurin for assistance to prepare the manuscript. The study was supported by the New Enterprise Incentive Scheme, an Australian Government initiative.

**Conflicts of Interest:** The author declares no conflict of interest.

## References

1. D.M. Higgins, R.J. Basaraba, A.C. Hohnbaum, E.J. Lee, D.W. Grainger, M. Gonzalez-Juarrero, Localized Immunosuppressive Environment in the Foreign Body Response to Implanted Biomaterials, The American Journal of Pathology, Vol. 175, No. 1, July 2009, 161-170.
2. S. Franz, S. Rammelt, D. Scharnweber, J.C. Simon, Immune responses to implants e A review of the implications for the design of immunomodulatory biomaterials, Biomaterials 32 (2011) 6692-6709.
3. J.M. Morais, F. Papadimitrakopoulos, D.J. Burgess, Biomaterials/Tissue Interactions: Possible Solutions to Overcome Foreign Body Response, AAPS J. 2010 Jun; 12(2): 188–196.
4. M. Kastellorizios, F. Papadimitrakopoulos, D.J. Burgess, Prevention of foreign body reaction in a pre-clinical large animal model, Journal of Controlled Release 202 (2015) 101–107.
5. J.E. Puskas, M.T. Luebbers, Breast implants: the good, the bad and the ugly. Can nanotechnology improve implants? WIREs Nanomed Nanobiotechnol, 2011. doi: 10.1002/wnan.164.
6. W.K. Ward, A Review of the Foreign-body Response to Subcutaneously-implanted Devices: The Role of Macrophages and Cytokines in Biofouling and Fibrosis, J Diabetes Sci Technol, 2(5), 2008, 768-777.
7. J.M. Anderson, A. Rodriguez, D.T. Chang, Foreign Body Reaction to Biomaterials, Semin Immunol. 2008 April ; 20(2): 86–100.
8. Begishev V ., Gavrilov N., Mesyats G., Klyachkin Y u., Kondyurina I., Kondyurin A., Osorgina I., Modification of polyurethane endoprosthetics surface by pulse ion beam, Proc. of 12th Intern. Conf. on High-Power Particle Beams, Haifa, Israel, June7-12, 1998. Ed. by M.Markovits and J. Shiloh, vol 2, p. 997-1000.
9. Mesyats G., Klyachkin Yu., Gavrilov N., Kondyurin A., Adhesion of Polytetrafluorethylene modified by an ion beam, Vacuum, vol.52, 1999, P. 285-289.
10. Kondyurin A., Maitz M.F., Surface Modification of ePTFE and Implants using the same, US patent WO 2007/022174 A3, 2007.
11. I. V. Kondyurina, V. S. Chudinov, V. N. Terpugov, A. V. Kondyurin, Influence of the Young's Modulus of Polyurethane Implants on the Organism's Immune Response, Biomedical Engineering, Vol. 52, No. 6, March, 2019, pp. 431434.
12. B.K. Gan, M.M.M. Bilek, A. Kondyurin, K. Mizuno, D.R. McKenzie, Etching and structural changes in nitrogen plasma immersion ion implanted polystyrene films, Nuclear Instruments and Methods in Physics Research B 247 (2006) 254–260.
13. K.S. Suvarna, C. Layton, J.D. Bancroft, Bancroft's Theory and Practice of Histological Techniques, 8th Edition, Elsevier Health Sciences, London, 2018, 584 pages.
14. A.C. Ferrari, J. Robertson, Interpretation of Raman spectra of disordered and amorphous carbon, Phys. Rev. 61 (2000) 14095.
15. N. Emanuel, A. Buchachenko, Chemical Physics of Polymer Degradation and Stabilization, VNU Science Press, Utrecht, 1987.

16. A.J.T. Teo, A. Mishra, I. Park, Y.-J. Kim, W.-T. Park, Y.-J. Yoon, Polymeric Biomaterials for Medical Implants and Devices, ACS Biomater. Sci. Eng., 2, 2016, 454–472.
17. Yung LY, Lim F, Khan MM, Kunapuli SP, Rick L, Colman RW, Cooper SL. Neutrophil adhesion on surfaces preadsorbed with high molecular weight kininogen under well-defined flow conditions. Immunopharmacology, 32 (1-3), 1996, 19-23.
18. Yung LY, RW Colman, Cooper SL. Neutrophil adhesion on polyurethanes preadsorbed with high molecular weight kininogen. Blood, 94(8), 1999, 2716-2724.
19. Mazaheri MK, Schultz GS, Blalock TD, Caffee HH, Chin GA. Role of connective tissue growth factor in breast implant elastomer capsular formation. Ann Plast Surg. 50(3), 2003, 263-268.
20. Miller M, Cho JY, McElwain K, McElwain S, Shim JY, Manni M, Baek JS, Broide DH. Corticosteroids prevent myofibroblast accumulation and airway remodeling in mice. Am J Physiol Lung Cell Mol Physiol., 290(1), 2006, L162-169.
21. M.N. Avula, A.N. Rao, L.D. McGill, D.W. Grainger, F. Solzbacher, Foreign body response to subcutaneous biomaterial implants in a mast cell-deficient Kitw-Sh murine model, Acta Biomaterialia 10 (2014) 1856–1863.
22. C. Hillgruber, A.K. Steingraber, B. Poppelmann, C.V. Denis, J. Ware, D. Vestweber, B. Nieswandt, S.W. Schneider, T. Goerge, Blocking Von Willebrand Factor for Treatment of Cutaneous Inflammation, Journal of Investigative Dermatology (2014) 134, 77–86.
23. B. Horvath, D. Hegedus, L. Szapary, Z. Marton, T. Alexy, K. Koltai, L. Czopf, I. Wittmann, I. Juricskay, K. Toth, G. Kesmarky, Measurement of von Willebrand factor as the marker of endothelial dysfunction in vascular diseases, Exp Clin Cardiol. 2004 Spring; 9(1): 31–34.
24. K.M. Provchy, Von Willebrand Factor Expression in Vascular Endothelial Cells of Cage Control and Antiorthostatic Cage Suspension Golden Hamster Ovaries, Senior Honors Thesis, East Tennessee State University, 2010.
25. L. Sheng, Q. Yu, F. Xie, Q. Li, Foreign body response induced by tissue expander implantation, Molecular Medicine Reports, 9 (2014) 872-876.
26. W.-J. Hu, J.W. Eaton, T.P. Ugarova, L. Tang, Molecular basis of biomaterial-mediated foreign body reactions, Blood, 98 (2001) 1231-1238.
27. 10. J.A. Jones, D.T. Chang, H. Meyerson, E. Colton, I.K. Kwon, T. Matsuda, J.M. Anderson, Proteomic analysis and quantification of cytokines and chemokines from biomaterial surface-adherent macrophages and foreign body giant cells. J Biomed Mater Res A. 83 (2007) 585-596.
28. V. Melnig, N. Apetroaei, N. Dumitrascu, Y. Suzuki, V. Tura, Improvement of polyurethane surface biocompatibility by plasma and ion beam techniques, Journal of Optoelectronics and Advanced Materials, 7(5), 2005, 2521 – 2528.
29. M. Iwaki, Ion surface treatments on organic materials, Nuclear Instruments and Methods in Physics Research, B 175-177, 2001, 368-374.
30. H. Chen, L. Yuan, W. Song, Z. Wu, D. Li, Biocompatible polymer materials: Role of protein–surface interactions, Progress in Polymer Science, 33, 2008, 1059–1087.
31. A. Krishnan, Y.-H. Liu, P. Cha, D. Allara, E.A. Vogler, Interfacial energetics of globular–blood protein adsorption to a hydrophobic interface from aqueous-buffer solution, J. R. Soc. Interface, 3, 2006, 283–301.
32. M. Bellion, L. Santen, H. Mantz, H. Haehl, A. Quinn, A. Nagel, C. Gilow, C. Weitenberg, Y. Schmitt, K. Jacobs, Protein adsorption on tailored substrates: long-range forces and conformational changes, J. Phys.: Condens. Matter, 20, 2008, 404226 (11pp).
33. Encyclopedia of Polymer Science and Technology, Ed. By Herman, Wiley, New York, 2004.
34. A. Kondyurin, M. Bilek, Ion Beam Treatment of Polymers. Application aspects from medicine to space, Second Edition, Elsevier, Oxford, 2014.
35. A. Kondyurin, N.J. Nosworthy, M.M.M. Bilek, R. Jones, P.J. Pigram, Surface Attachment of Horseradish Peroxidase to Nylon Modified by Plasma-Immersion Ion Implantation, Journal of Applied Polymer Science, 120, 2011, 2891–2903.
36. A. Kondyurin, N.J. Nosworthy, M.M.M. Bilek, Effect of Low Molecular Weight Additives on Immobilization Strength, Activity, and Conformation of Protein Immobilized on PVC and UHMWPE, Langmuir, 27, 2011, 6138–6148.

37. A.V. Kondyurin, P. Naseri, J.M.R. Tilley, N.J. Nosworthy, M.M.M. Bilek, D.R. McKenzie, Mechanisms for Covalent Immobilization of Horseradish Peroxidase on Ion-Beam-Treated Polyethylene, Scientifica, 2012, Article ID 126170, 28 pages, http://dx.doi.org/10.6064/2012/126170.
38. A. Kondyurin, N.J. Nosworthy, M.M.M. Bilek, Attachment of horseradish peroxidase to polytetrafluorethylene (teflon) after plasma immersion ion implantation, Acta Biomaterialia, 4, 2008, 1218–1225.
39. Kiaei D, Hoffman AS, Horbett TA, Tight binding of albumin to glow discharge treated polymers. J Biomater Sci Polym Ed, 4, 1992, 35–44.
40. W. Khan, M. Kapoor, N. Kumar, Covalent attachment of proteins to functionalized polypyrrole-coated metallic surfaces for improved biocompatibility, Acta Biomaterialia, 3, 2007, 541–549.
41. C. Mateo, V. Grazu, J. M Palomo, F. Lopez-Gallego, R. Fernandez-Lafuente, J.M. Guisan, Immobilization of enzymes on heterofunctional epoxy supports, Nature Protocols, 2(5), 2007, 1023.
42. C.D. Hodneland, Y.-S. Lee, D.-H. Min, M. Mrksich, Selective immobilization of proteins to self-assembled monolayers presenting active site-directed capture ligands, PNAS, 99, 2002, 5048-5052.
43. B. K. Gan, A. Kondyurin, M.M.M. Bilek, Comparison of Protein Surface Attachment on Untreated and Plasma Immersion Ion Implantation Treated Polystyrene: Protein Islands and Carpet, Langmuir, 23, 2007, 2741-2746.
44. Hirsh, S., McKenzie, D., Nosworthy, N., Denman, J., Sezerman, O., Bilek, M., The Vroman effect: Competitive protein exchange with dynamic multilayer protein aggregates. Colloids And Surfaces B: Biointerfaces, 103, 2013, 395- 404.
45. S. E. Stein, R. L. Brown, Chemical theory of raphite-like molecules, Carbon, 23, 1985, 105-109. 28.
46. V. Svorcik, V. Hnatowicz, P. Stopka, L. Bacakova, J. Heitze, R. Ochsner, H. Ryssel, Amino acids grafting of Ar+ ions modified PE, Radiation Physics and Chemistry, 60, 2001, 89–93.
47. M.J. Davies, The oxidative environment and protein damage, Biochimica et Biophysica Acta, 1703, 2005, 93–109.
48. M.S. Lord, W. Yu, B. Cheng, A. Simmons, L. Poole-Warren, J. M. Whitelock, The modulation of platelet and endothelial cell adhesion to vascular graft materials by perlecan, Biomaterials, 30, 2009, 4898–4906.
49. J.H. Pang, Y. Farhatnia, F. Godarzi, A. Tan, J. Rajadas, B.G. Cousins, A.M. Seifalian, In situ Endothelialization: Bioengineering Considerations to Translation, Small, 11 (47), 2015, 6248–6264.
50. J.G. Steele, T.A. Savolainen, A. Dalton, L. Smith, G.J. Smith, Adhesion to Laminin and Expression of Laminin in Clonally Related Transformed and Control Sublines from an Alveolar Epithelial Cell Strain, Cancer research, 50, 1990, 3381-3389.
51. V. S. Chudinov, I. V. Kondyurina, I.N. Shardakov, A. L. Svistkov, I. V. Osorgina, A. V. Kondyurin, Polyurethane Modified with Plasma-Ion Implantation for Medical Applications, Biophysics, 2018, Vol. 63, No. 3, pp. 330–339.
52. J.-S. Lee, M. Kaibara, M. Iwaki, H. Sasabe, Y. Suzuki, M. Kusakabe, Selective adhesion and proliferation of cells on ion-implanted polymer domains, Biomaterials, 14(12), 1993, 958.
53. B. Pignataro, E. Conte, A. Scandurra, G. Marletta, Improved cell adhesion to ion beam-irradiated polymer surfaces, Biomoterials, 18, 1997, 1461-1470.
54. N. Ozkucur, E. Richter, C. Wetzel, R.H.W. Funk, T.K. Monsees, Biological relevance of ion energy in performance of human endothelial cells on ion-implanted flexible polyurethane surfaces, J Biomed Mater Res, 93A, 2010, 258–268.
55. M.Y. Alkawareek, Q.Th. Algwari, G. Laverty, S.P. Gorman, W.G. Graham, D. O'Connell, B.F. Gilmore, Eradication of Pseudomonas aeruginosa Biofilms by Atmospheric Pressure Non-Thermal Plasma, PLoS ONE 7(8): e44289. doi:10.1371/journal.pone.0044289.